\documentclass[letter]{aa}
\usepackage{graphicx}
\usepackage[intlimits,sumlimits]{amsmath}
\usepackage{deluxetable}
\usepackage{times,epsfig} 
\usepackage{natbib}
\usepackage{amssymb}
\bibpunct{(}{)}{;}{a}{}{,} 
\usepackage{amsmath}
\usepackage[english]{babel}

\newcommand{\beqa}{\begin{eqnarray}} 
\newcommand{\eeqa}{\end{eqnarray}}

\newcommand{\bsub}{\begin{subequations}}
\newcommand{\esub}{\end{subequations}}
\newcommand{\beal}{\begin{align}}
\newcommand{\ealn}{\end{align}}

\begin{document}
\title{Measuring the dark matter velocity anisotropy in galaxy clusters}

\titlerunning{Measuring velocity anisotropy}
\authorrunning{Hansen \and Piffaretti}
\author{Steen H. Hansen \inst{1} \and Rocco Piffaretti \inst{2}}

\institute{Dark Cosmology Centre, Niels Bohr
Institute, University of Copenhagen, Juliane Maries Vej 30,
Copenhagen, Denmark
\and
SISSA/ISAS, via Beirut 4, 34014 Trieste, Italy} 

\offprints{S. H. Hansen, hansen@dark-cosmology.dk}
\date{Received May 22, 2007 / Accepted ??}
\abstract
{}{The Universe contains approximately 6 times more dark matter
than normal baryonic matter, and a directly observed
fundamental difference between dark matter and baryons would both
be significant for our understanding of dark matter
structures and provide us with information about the basic
characteristics of the dark matter particle.
We discuss one
distinctive feature of  dark matter structures in equilibrium, namely the
property that a local dark matter temperature may depend on
direction. This is in stark contrast to
baryonic gases.}
{We used X-ray observations of two nearby, relaxed
galaxy clusters, under the assumptions of hydrostatic equilibrium
and identical dark matter and gas temperatures in the outer cluster region,
to measure this dark matter temperature anisotropy
$\beta_{\rm dm}$, with non-parametric Monte Carlo methods.}
{We find
that $\beta_{\rm dm}$ is greater than the value predicted for baryonic
gases, $\beta_{\rm gas}=0$, at more than $3\sigma$ confidence.  The
observed value of the temperature anisotropy is in fair agreement with
the results of cosmological N-body simulations and shows that the
equilibration of the dark matter particles is not governed by local
point-like interactions in contrast to baryonic gases.}

\keywords{Cosmology: dark matter --- X-rays: galaxies: clusters  --- 
Galaxies: clusters: individual: A2052, Sersic 159-03
\maketitle
\section{Introduction}


The possibility that the temperature of the dark matter depends on direction is
usually expressed through the velocity aniso\-tro\-py
\begin{equation}
\beta_{\rm dm} \equiv 1 - \frac{\sigma^2_t}{\sigma^2_r} ~,
\end{equation}
where $\sigma^2_t$ and $\sigma^2_r$ are the 1-dimensional tangential and radial
velocity dispersions in a spherical system~\citep{binneytremaine}. Our intuition from
classical gases leads us to loosely refer to these dispersions as the dark matter
temperatures in the tangential and radial directions with respect to
the centre of the galaxy cluster.  If most dark matter particles in an
equilibrated structure were purely on radial orbits, then $\beta_{\rm
dm}$ could be as large as 1 and, for mainly tangential orbits
$\beta_{\rm dm}$ could be arbitrarily large and negative.

The tempe\-ra\-ture of a baryonic gas is only well-defined when the
gas is locally in thermal equilibrium, and this gas equilibration is
achieved through point-like interactions. In that case the gas
temperature is independent of direction, which is expressed as
$\beta_{{\rm gas}} = 0$.  
This is because the mean free path, typically in the tens of kpc
\citep{sarazin}, is much shorter than the cluster scale of Mpc. 
If dark matter was collisional and hence
achieved equilibration through collisions, then one would also have
$\beta_{\rm dm}=0$.

Numerical N-body simulations of collision-less dark matter particles
show that the dark matter temperature aniso\-tro\-py is different from zero,
and for galaxy clusters these simulations show that $\beta_{\rm dm}$ goes from
zero in the central region to 0.4-0.6 towards the outer
region~\citep{colelacey,carlberg,hansenmoore,hansenstadel}. 
A mass-averaged $\beta_{\rm dm}$
is close to 0.3.

During the assembly of galaxy clusters, the baryonic gas is
shock-heated and eventually achieves energy equipartition with the
gravitationally dominating dark matter at a temperature that is
directly related to the temperature of the dark matter. 
This is true as long as radiation and similar non-gravitational
effects are negligible.
The most
sensible  definition of dark
matter temperature is by averaging over the 3 directions
\begin{equation}
T_{\rm dm} \, \frac{k_B}{\mu m_p} \equiv \frac{1}{3} \left( \sigma_r^2 + 2 \sigma_t^2 \right) 
= \sigma^2_r \left(1 - \frac{2}{3}\beta_{\rm dm} \right) 
\label{eq:temp}
\end{equation}
and we describe in the next section how to derive this dark
matter temperature from X-ray observations of the gas.  
Here  $k_B$ is the Boltzmann constant, $m_p$ the proton mass, 
and $\mu$ the mean molecular weight (we assume $\mu=0.61$). 
We later
discuss to what extent the assumption of equality between gas and dark matter
temperatures is supported by numerical simulations (from an average over a large
set of simulated clusters, some of which are significantly more perturbed
than the ones considered here), and
estimate the possible effect on $\beta_{\rm dm}$. 

Our approach does not
make any assumptions about the parametric form of mass or dispersion
profiles in contrast to earlier related studies~\citep{priya,ikebe}.
We later demand that the
reconstructed dark matter temperature and the observed gas temperature
must be equal in equilibrated regions that are not affected by
radiative processes. This is a fair assumption since it only relies on
the principle of equipartition between the dark matter and the
gas. However, we emphasise that this is an assumption to be
tested in the near future on high-resolution numerical simulations.

\section{Finding the dark matter temperature}

For relaxed and spherically symmetric galaxy clusters, one can use X-ray observations of
the hot, ionized gas to deduce the de-projected gas temperature and
gas density as functions of radius.  These are needed in the equation of
hydrostatic equilibrium~\citep{fabricant,sarazin}, which for spherical structures
relates the total mass of gra\-vi\-tating matter at a given radius to the
radial dependence of gas temperature and density
\begin{equation}
M(r) = - \frac{k_B T_e(r) \, r}{\mu m_pG} \left( \frac{d{\rm ln} n_e}{d {\rm ln} r}
+ \frac{d{\rm ln} T_e}{d {\rm ln} r} \right) ~ ,
\label{eq:hydr}
\end{equation}
where $G$ is the gravitational constant.
We here
assume that turbulence is negligible, which will have to be tested
in the future e.g.\ by using line-broadening in
metal lines~\citep{sunyaev}. 

We consider two highly relaxed
clusters, which are likely to have only very little turbulence. 
That equilibrated
structures have reliably reconstructed mass (i.e. obey the hydrostatic
equilibrium) was supported in the comparison between lensing and X-ray
observations \citep{allen98}. We
also assume that non-gravitational entropy injection into the baryonic
gas (e.g. from a central AGN) can be ignored in the region we
consider. Radio cavities are typically located within much smaller cluster-centric
distances  than those excised in our analysis
\citep[e.g.][]{birzan}
We thus have both the total mass and the baryonic mass (since the
cluster plasma is optically thin); since the stellar component is
negligible~\citep{1998ApJ...503..518F}, we can easily get the dark
matter mass and dark matter density as functions of radius.

To make the connection with the dark matter temperature, we
must consider the Jeans equation~\citep{binneytremaine}, which relates
the dark matter density and velocity dispersions with the total
gravitating mass:
\begin{equation}
M(r) = -\frac{\sigma^2_r \, r}{G} \left( \frac{d {\rm ln} \rho}{d {\rm ln} r} +
\frac{d {\rm ln} \sigma^2_r}{d {\rm ln} r} + 2 \beta_{\rm dm} \right) ~.
\label{eq:jeans}
\end{equation}
We are assuming that the Jeans equation for spherical structures
is accurate for dark matter, which has been shown to be a good assumption
by numerical
simulations~\citep{rasia}.
Equation~(\ref{eq:jeans}) can be integrated to give 
\begin{equation}
\sigma_r^2 (R) = \frac{G}{\tilde \rho (R)} 
\int _R^\infty \frac{M(r) \tilde \rho(r)}{r^2} dr ~,
\label{eq:sigma}
\end{equation}
where $\tilde \rho$ is defined from the dark matter density and anisotropy
\begin{equation}
\frac{d {\rm ln} \tilde \rho}{d {\rm ln } r}
= \frac{d {\rm ln} \rho_{\rm dm}}{d {\rm ln } r} + 2 \beta_{\rm dm} ~.
\label{eq:betatilde}
\end{equation}
Equations.~(\ref{eq:hydr}-\ref{eq:betatilde})
provide all the tools needed to use the observed gas density and
gas temperature to derive the dark matter temperature.  
The only free parameter
is the dark matter anisotropy, $\beta_{\rm dm}$. 
We assume for simplicity that
$\beta_{\rm dm}$ can be treated as a constant throughout the observed
part of the galaxy cluster, a simplification that future data clearly
will be able to lift.
Thus, for any given value of $\beta_{\rm dm}$, we can calculate the
DM temperature, which can then be compared with the gas temperature. This
gives us the possibility of comparing different values of $\beta_{\rm dm}$ 
by standard statistical means.

\section{Two quiet clusters}

We used X-ray data from {\it XMM-Newton} of the two nearby galaxy
clusters A~2052 and S\'ersic~159$-$03 (also known as Abell S1101).  
These are highly relaxed
clusters, where a temperature decrement in the central region is
clearly identified in the de-projected data~\citep{kaastra,rocco}.
These two clusters were chosen for three reasons. First of all, 
to extract the dark matter temperature non-parametrically one
needs data at large radii. Secondly, the cluster surface-brightness
map must be highly circular, with no evidence of a substructure, and
the visual inspection of the de-projected temperature must show a
smooth behaviour, indicating a fully relaxed cluster. Finally, the
de-projected gas temperature must be robustly observed. These
clusters were analysed assuming the standard cosmological $\Lambda$CDM
model by \cite{rocco}, who show that the outer temperature decreases
by approximately $30\%$ from the maximum temperature.  These authors
also show that a one-component gas temperature gives an excellent
fit to the observed spectra. Even though the gas temperature profiles of
these two clusters are slightly different, they both appear to
provide a fair fit to a universal temperature profile~\citep{rocco}. 

\section{Non-parametric analysis}

We treat the data in an entirely non-parametric way by Monte
Carlo methods.  The input data is a set of $7$ radial values for the
de-projected gas temperature and density and their corresponding
error bars. 
These are determined from de-projected, spatially resolved spectra. A
detailed description of the de-projection technique is given in \cite{kaastra}. 

For each radial bin, we
select randomly a temperature and density with a Gaussian-distributed
value around the observed number and a width corresponding to the
observed error bars. We then use the equation of hydrostatic
equilibrium (Eq.~\ref{eq:hydr}) to derive the total mass as function of radius.  This
total mass extends only to a radius of 0.5-0.8 Mpc,
so we select a random number between $-4$ and $-2$ for the logarithmic
dark matter density slope at larger radii, which is a larger range
than expectations from both simulation and theory~\citep[e.g.][]{diemand}. The
gas mass is already 
negligible beyond the outermost bin.
We checked that varying these assumptions has
virtually no influence on the results.

We can now subtract the gas mass from the total mass.
Having both the total mass and the dark matter density as functions of
radius, as well as the assumed value for $\beta_{\rm dm}$, allows us
to integrate eq.~(\ref{eq:sigma}) and hence get the dark matter
temperature from eq.~(\ref{eq:temp}). At no point do we make any
assumption about the form of the dark matter or gas profiles, nor
about boundary conditions. This is in contrast to earlier related
studies~\citep{priya,ikebe}.

For each radial bin we now generate $10\,000$ models, and we can proceed with
a frequentists statistical analysis.  We calculate the median value of
all the derived model temperatures, and we select the range covered by
the central $70-75\%$ of these models as representative of the
error bars of the reconstructed dark matter temperature, $\Delta
T_{\rm dm}$.  This number of central models is chosen to make the best
$\chi ^2$ per degree of freedom of order unity. We are being
conservative since we are including more than the normal $68\%$ of the
models, corresponding to 1 standard deviation for a gaussian 
probability distribution.  As
total error bar we use the quadratic sum of this reconstructed error and the
observed temperature error bars, $\Delta T_e$ (which is anyway much smaller). 
It is worthwhile to mention that a few of the produced models
are non-physical in the sense that
they may have decreasing mass locally; however, we choose the most
conservative approach and keep all models, hence slightly increasing
the error bars while only very mildly shifting the median.

In Fig.~1 we show the
observed gas and reconstructed dark matter temperatures for A~2052 (top-panel)
and S\'ersic~159$-$03 (bottom panel),
with their corresponding error bars, for the case of $\beta_{\rm
dm}=0.6$. 
We see from Fig.~1 that the gas and dark matter
temperatures are in good agreement in the outer region.
This value of $\beta$ was chosen because it leads to reasonable
agreement between the dark matter and gas temperatures. For a different
value of $\beta$, the reconstructed DM temperature will be different
according to Eqs.~(\ref{eq:temp}, \ref{eq:sigma}). The main effect comes
from Eq.~\ref{eq:temp} and shows that a smaller (or negative) $\beta$
will imply a higher reconstructed temperature.

\begin{figure}[thb]
	\centering
	\includegraphics[angle=0,width=0.49\textwidth]{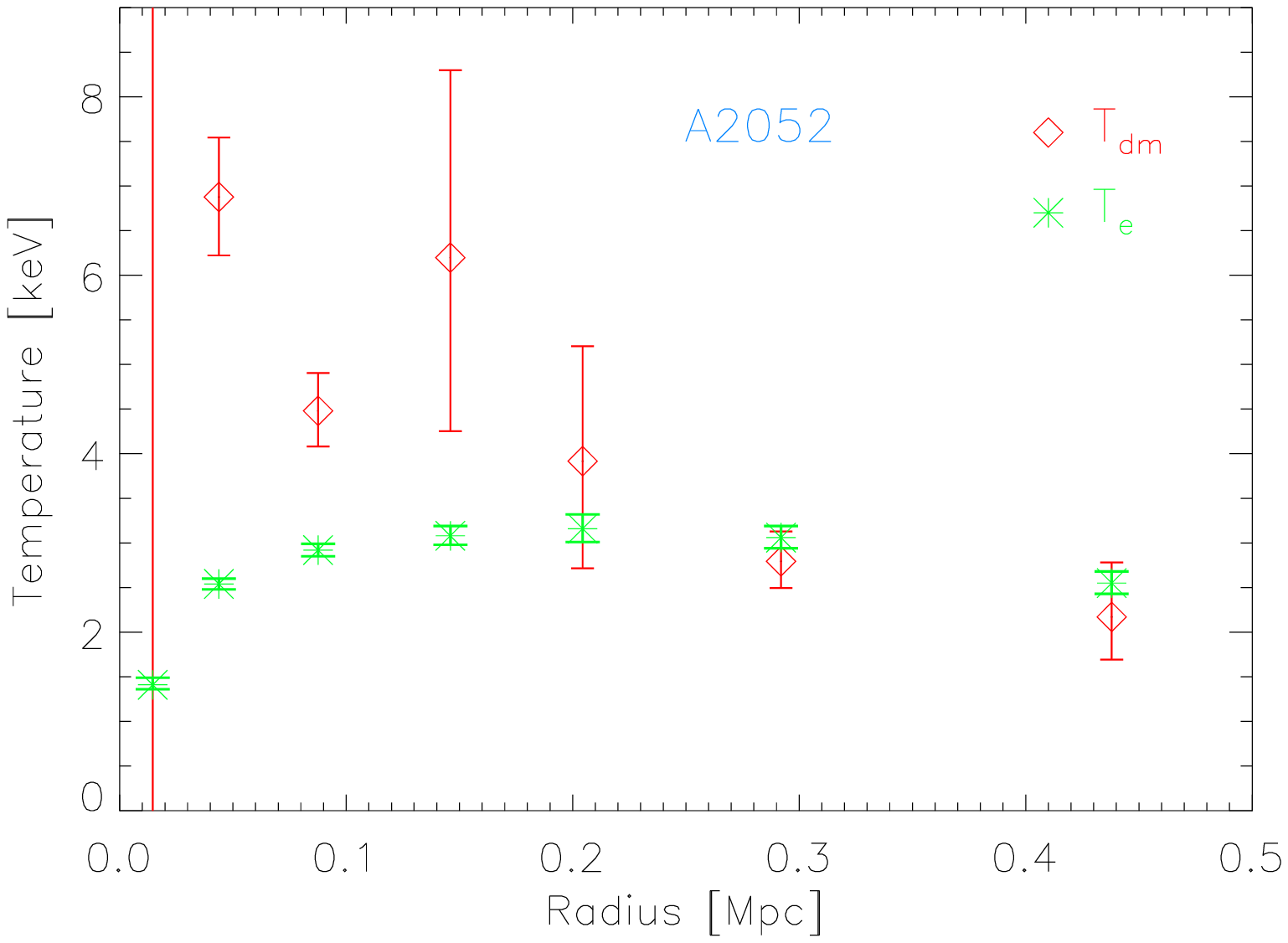}
	\includegraphics[angle=0,width=0.49\textwidth]{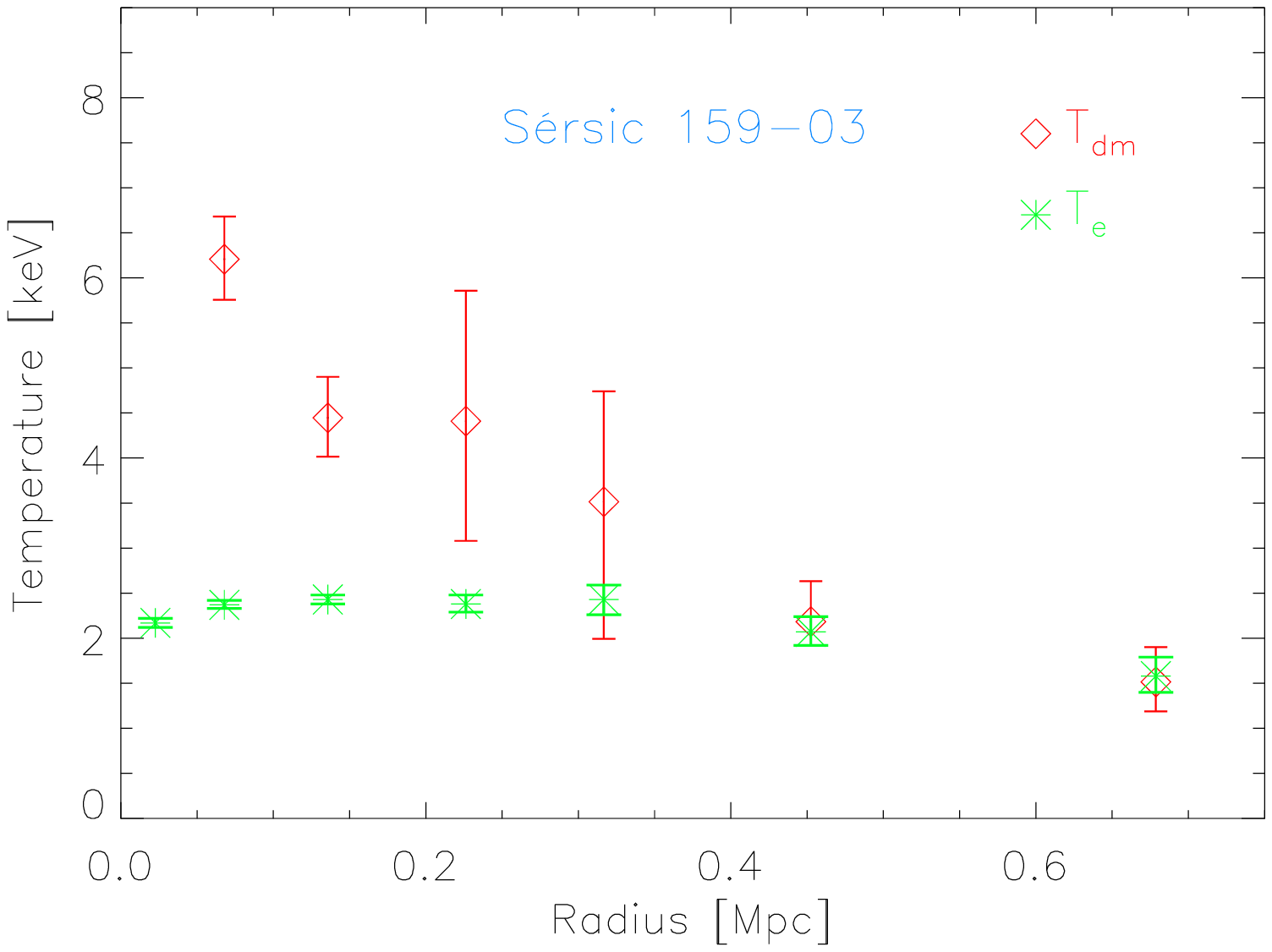}
	\caption{Observed baryonic (green stars) and reconstructed
dark matter (red diamonds) temperatures with error bars for A~2052
(top) and S\'ersic~159$-$03 (bottom), as function of radius. A dark
matter velocity anisotropy of $\beta_{\rm dm}=0.6$ is assumed in both plots,
because it gives a reasonable fit to the outer
region. The radiative cooling is prominent in the central dense
region, and in the analysis we include only the 4 outermost
temperature bins.}
\label{fig1}
\end{figure}

There are large radial variations due to the non-parametric treatment: small
non-monotonous variations in the measured gas temperature and density imply
non-trivial variation in the reconstructed dark matter temperature. Similarly,
the error bars of the reconstructed dark matter temperature directly reflect
the error bars of the measured gas density and especially temperature.

The central region of the cluster is very dense, and the gas
temperature is most likely governed by radiative and conductive
processes~\citep[e.g.][]{sarazin}, so we expect the dark matter and gas
temperatures to agree only in the outer part. 
The corresponding radius where
radiative processes become important is roughly where the temperature
starts decreasing as one approaches the centre. 
In particular, the
central dark matter temperatures appear to have a quite different radial behaviour
from the shape expected from numerical
simulations~\citep[e.g.][]{eke,rasia}. This is clearly related to these being
derived directly from the gas profile, which is strongly affected in the
central region  by non-gravitational processes.
The most recent
numerical simulations show good agreement between the simulated and
observed gas temperatures in the outer region~\citep{pratt07}, whereas
the central region still allows for improvements in the simulations
to reach agreement with observations.

We have discussed how to reconstruct the dark matter temperature for a
given assumed velocity anisotropy above. We therefore proceed and reconstruct
the DM temperature for a range of different velocity anisotropies, and then
compare them.

We assume that gas and DM temperatures in the 4 outer-most radial bins must
agree are do thus not include the region dominated by radiative
processes. This allows us to perform a $\chi ^2$ comparison between
the reconstructed model and the observed temperature. We show the
resulting figure of $\Delta \chi ^2$ as a function of $\beta_{\rm dm}$
in Fig.~2. One finds that a positive and non-zero dark matter
anisotropy is preferred, with $\beta_{\rm dm} > 0.2$ for
S\'ersic~159$-$03 (and $\beta_{\rm dm} > 0$ for A~2052) at $\Delta \chi
^2 =9$, corresponding to $3\sigma$ confidence.  The two structures
have $\beta_{\rm dm} < 1$ only at $\Delta \chi ^2 =3 (7)$. This is the
first time the dark matter temperature anisotropy has been measured,
and it is comforting that it agrees fairly well with cosmological N-body
simulations, which find $\beta_{\rm dm} \sim 0.3$. We emphasise that we
do not claim that the clusters have anisotropies of 0.45 or 0.7
(which naturally would disagree with numerical simulations), but
are merely stating that the anisotropies are greater than zero and 
consistent with numerical simulations.

\begin{figure}[thb]
	\centering
	\includegraphics[angle=0,width=0.49\textwidth]{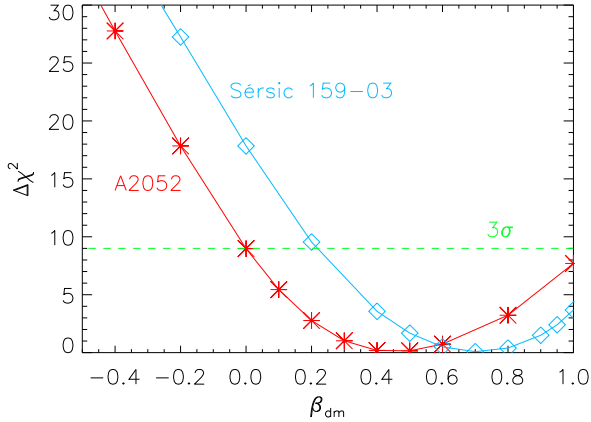}
	\caption{$\Delta \chi ^2$ as a function of $\beta_{\rm
dm}$. Both galaxy clusters have a preference for radial dark
matter velocity anisotropy, with best-fit values at 0.45 (A~2052, red
stars) and 0.7 (S\'ersic~159$-$03, blue diamonds). The $3\sigma$
confidence level is shown with the horizontal (green dashed) line.  A
vanishing $\beta_{\rm dm}$ is ruled out at $\Delta \chi ^2$ of 9 and
18.}
\label{fig:2}
\end{figure}

For robustness we checked that keeping $68\%$ of the models gives
$\chi ^2$ per degree of freedom of 1.5 for A~2052 (and 1.1 for
S\'ersic~159$-$03), while keeping the best-fit points for $\beta_{\rm
dm}$ virtually unchanged. In this case the resulting error bars of
$\beta_{\rm dm}$ become slightly more stringent.

We performed the same study using only the outermost 3 bins in the
statistical analysis, and the central values for $\beta_{\rm dm}$ were found to
be near 0.4 for A~2052 and at 0.7 for S\'ersic~159$-$03, in excellent agreement
with the results above. In this case one only keeps $40\%$ of the
models in order to get $\chi ^2$/d.o.f.\ close to unity, so we
do not trust the resulting (slightly more restrictive) confidence
interval for $\beta_{\rm dm}$ in that case.

\section{Discussion}
The selected clusters appear to be fully relaxed; however, as noted
above, the possibility of bulk motion in the
gas still exists. Such a bulk motion implies that the reconstructed total mass is
underestimated since there will then be an extra term on the r.h.s.\ of 
Eq.~(3). Numerical simulations, including both gas and DM, have shown that
for equilibrated structures this underestimation in the reconstructed
mass is less than $15\%$~\citep{rasia}. This implies that
$\sigma^2_r$ is underestimated by at most $15\%$ (Eq.~(5)). In
addition, the dark matter temperature will be underestimated, since the
energy of the gas also contains a term from the bulk motion.  This
underestimate is also at most $15\%$~\citep{rasia} in the
region considered. These two effects partially (possibly almost
totally) cancel out in the determination of the velocity anisotropy.
Therefore $\beta_{\rm dm}$ is accurately determined, and the error from a
possible bulk motion in the gas is significantly smaller than $20\%$
(see Eq.~(2))~\footnote{We are in the process of quantifying these
estimates through high resolution numerical simulations, and the
results will be presented elsewhere.}.

We have here considered two clusters that both have a central
temperature decrement, and we see that the reconstructed dark matter
temperature is very different from the gas temperature in the central
region. There is growing evidence that clusters separate into two
distinct classes, where one class has decreasing central
temperature, the so-called cool-core clusters like the two structures
considered here, and the second class has a roughly constant temperature
in the central region~\citep[e.g.][]{sanderson}.  It will be very interesting
to perform another study similar to the one made here, to see if
these non-cool-core clusters have dark matter temperatures that may
even agree in the central region.

It is worth mentioning that the temperature anisotropy can be
measured in principle in an underground directional sensitive detector; however, it
will require a large dedicated experimental programme~\citep{olehost}.

Studies of highly non-equilibrated merging
clusters~\citep[e.g.][]{clowe} have shown that the dark matter has
a much smaller scattering cross section than baryonic gas.  We have here
extended our understanding of dark matter to show that the
equilibration of dark matter structures is not governed by point-like
collisions. This demonstrates that dark matter behaviour is fundamentally
different from baryons.  This is particularly important for
understanding what drives structure formation
and the evolution of cosmological structures. The standard theory of 
structure formation is
based on the one basic assumption that dark matter is indeed
collisionless, and we here provide observational evidence that
this is a correct assumption.

\begin{acknowledgements} 
  It is a great pleasure to thank J. Hjorth, O. Host, and K. Pedersen
  for discussions, and Gary Mamon for constructive
  suggestions.  The Dark Cosmology Centre is funded by the Danish
  National Research Foundation.
\end{acknowledgements}


\end{document}